\documentclass[prl, reprint,showpacs,  superscriptaddress]{revtex4-1}

\usepackage{color,xcolor}

\usepackage{footnote}
\usepackage[colorinlistoftodos,bordercolor=red!50,backgroundcolor=red!10,linecolor=red!50,textsize=tiny]{todonotes}
\usepackage{changes}
\usepackage[utf8]{inputenc}
\usepackage[T1]{fontenc}
\usepackage[english]{babel}  % English language
\usepackage{times}
\usepackage{bm}
\usepackage{ulem}

\usepackage[colorlinks=true,linkcolor=red,citecolor=red,urlcolor=red,pdfauthor={ },pdftitle={ },pdfsubject={ },pdfkeywords={ }]{hyperref}
\usepackage{amssymb,amsmath,amsfonts,amsthm}
\usepackage{mathtools}
\usepackage{verbatim}
\usepackage{enumerate}
\usepackage{graphicx}
\usepackage{hyperref}
\usepackage{bbm}
\usepackage{booktabs}

\usepackage[caption=false]{subfig}

\usepackage{setspace}
\usepackage{array}

\usepackage{color}
\definecolor{dred}{rgb}{.8,0.2,.2}
\definecolor{ddred}{rgb}{.8,0.5,.5}
\definecolor{dblue}{rgb}{.2,0.2,.8}
\definecolor{dgreen}{rgb}{.2,0.5,.2}

\newcommand{\bra}[1]{\mbox{$\langle #1|$}}
\newcommand{\ket}[1]{\ensuremath{|#1\rangle}}

\newcommand{\be}{\begin{equation}}
\newcommand{\ee}{\end{equation}}
\newcommand{\bea}{\begin{eqnarray}}
\newcommand{\eea}{\end{eqnarray}}

\begin{document}

\title{Measuring Quantum Entanglement from Local Information by Machine Learning}

\author{Yulei Huang}
\affiliation{Shenzhen Institute for Quantum Science and Engineering and Department of Physics, Southern University of Science and Technology, Shenzhen 518055, China}
\affiliation{Guangdong Provincial Key Laboratory of Quantum Science and Engineering,
Southern University of Science and Technology, Shenzhen 518055, China}
\affiliation{Shenzhen Key Laboratory of Quantum Science and Engineering, Southern University of Science and Technology, Shenzhen,518055, China}

\author{Liangyu Che}
\affiliation{Shenzhen Institute for Quantum Science and Engineering and Department of Physics, Southern University of Science and Technology, Shenzhen 518055, China}
\affiliation{Guangdong Provincial Key Laboratory of Quantum Science and Engineering,
Southern University of Science and Technology, Shenzhen 518055, China}
\affiliation{Shenzhen Key Laboratory of Quantum Science and Engineering, Southern University of Science and Technology, Shenzhen,518055, China}

\author{Chao Wei}
\affiliation{Shenzhen Institute for Quantum Science and Engineering and Department of Physics, Southern University of Science and Technology, Shenzhen 518055, China}
\affiliation{Guangdong Provincial Key Laboratory of Quantum Science and Engineering,
Southern University of Science and Technology, Shenzhen 518055, China}
\affiliation{Shenzhen Key Laboratory of Quantum Science and Engineering, Southern University of Science and Technology, Shenzhen,518055, China}

\author{Feng Xu}
\affiliation{Shenzhen Institute for Quantum Science and Engineering and Department of Physics, Southern University of Science and Technology, Shenzhen 518055, China}
\affiliation{Guangdong Provincial Key Laboratory of Quantum Science and Engineering,
Southern University of Science and Technology, Shenzhen 518055, China}
\affiliation{Shenzhen Key Laboratory of Quantum Science and Engineering, Southern University of Science and Technology, Shenzhen,518055, China}

\author{Xinfang Nie}
\affiliation{Shenzhen Institute for Quantum Science and Engineering and Department of Physics, Southern University of Science and Technology, Shenzhen 518055, China}
\affiliation{Guangdong Provincial Key Laboratory of Quantum Science and Engineering,
Southern University of Science and Technology, Shenzhen 518055, China}
\affiliation{Shenzhen Key Laboratory of Quantum Science and Engineering, Southern University of Science and Technology, Shenzhen,518055, China}

\author{Jun Li}
\email{lij3@sustech.edu.cn}
\affiliation{Shenzhen Institute for Quantum Science and Engineering and Department of Physics, Southern University of Science and Technology, Shenzhen 518055, China}
\affiliation{Guangdong Provincial Key Laboratory of Quantum Science and Engineering,
Southern University of Science and Technology, Shenzhen 518055, China}
\affiliation{Shenzhen Key Laboratory of Quantum Science and Engineering, Southern University of Science and Technology, Shenzhen,518055, China}

\author{Dawei Lu}
\email{ludw@sustech.edu.cn}
\affiliation{Shenzhen Institute for Quantum Science and Engineering and Department of Physics, Southern University of Science and Technology, Shenzhen 518055, China}
\affiliation{Guangdong Provincial Key Laboratory of Quantum Science and Engineering,
Southern University of Science and Technology, Shenzhen 518055, China}
\affiliation{Shenzhen Key Laboratory of Quantum Science and Engineering, Southern University of Science and Technology, Shenzhen,518055, China}

\author{Tao Xin}
\email{xint@sustech.edu.cn}
\affiliation{Shenzhen Institute for Quantum Science and Engineering and Department of Physics, Southern University of Science and Technology, Shenzhen 518055, China}
\affiliation{Guangdong Provincial Key Laboratory of Quantum Science and Engineering,
Southern University of Science and Technology, Shenzhen 518055, China}
\affiliation{Shenzhen Key Laboratory of Quantum Science and Engineering, Southern University of Science and Technology, Shenzhen,518055, China}

%%%%%%%%%%%%%%%%%%%%%%%%%%renyi+Pn+coherence(no negativity)
\begin{abstract}
Entanglement is a key property in the development of quantum technologies and in the study of quantum many-body simulations. However, entanglement measurement typically requires quantum full-state tomography (FST). Here we present a neural network-assisted protocol for measuring entanglement in equilibrium and non-equilibrium states of local Hamiltonians. Instead of FST, it can learn comprehensive entanglement quantities from single-qubit or two-qubit Pauli measurements, such as Rényi entropy, partially-transposed (PT) moments, and coherence.  It is also exciting that our neural network is able to learn the future entanglement dynamics using only single-qubit traces from the previous time. In addition, we perform experiments using a nuclear spin quantum processor and train an adoptive neural network to study entanglement in the ground and dynamical states of a one-dimensional spin chain. Quantum phase transitions (QPT) are revealed by measuring static entanglement in ground states, and the entanglement dynamics beyond measurement time is accurately estimated in dynamical states. These precise results validate our neural network. Our work will have a wide range of applications in quantum many-body systems, from quantum phase transitions to intriguing non-equilibrium phenomena such as quantum thermalization.
  \end{abstract}

\maketitle
%%%%%%%%%%%%%%%%%%%%%%%%%%%%%%%%%%%%%%%%%%%%%%%%%%%%%
\textit{Introduction.} -- 
 Entanglement plays a crucial role in the development of quantum technologies \cite{horodecki2009quantum}. It is an essential resource for simulating many-body physics \cite{RevModPhys.86.153,amico2008entanglement}, investigating quantum advantages in quantum computation, and ensuring the security of quantum communication \cite{preskill2018quantum, gisin2007quantum}. Unfortunately, entanglement measures are not physical observables, making detection and quantification extremely challenging \cite{lu2016tomography}.

In the Noisy Intermediate-Scale Quantum (NISQ) era, new demands for quantum entanglement detection are being put forth. \emph{Demand 1}. Quantifying entanglement effectively. Although a large number of entanglement witnesses have been constructed \cite{sperling2013multipartite,zwerger2019device,huber2014witnessing}, some of which only require partial system information \cite{frerot2022unveiling}, they only provide a yes or no answer as to whether entanglement exists or not. Accurate quantification of entanglement, such as logarithmic negativity \cite{plenio2005logarithmic} and the PT moments \cite{yu2021optimal, neven2021symmetry}, typically require quantum FST since there are no observables for them. However, due to the exponential increase of the number of measurements with system size, performing FST to measure entanglement will no longer be practical for NISQ devices. In recent years, a large number of methods have been proposed to boost the efficiency of FST \cite{cramer2010efficient,torlai2018neural,gross2010quantum,xin2017quantum,xin2019local,xin2020quantum}, including compressed sensing \cite{gross2010quantum}, FST via local measurements \cite{xin2017quantum,xin2019local}, and neural-network FST \cite{torlai2018neural},  but these methods may not be effective for entanglement measurement since two close states may have significantly different values of entanglement \cite{bengtsson2017geometry}. Therefore, it is increasingly essential to develop effective entanglement measurement techniques. There have been extensive studies aimed at meeting this need \cite{koutny2022deep,huang2020predicting,gray2018machine,brydges2019probing}. For example, proposed a machine learning-assisted approach to measure entanglement, where the PT moments are measured as input layers and the logarithmic negativity is predicted \cite{gray2018machine}. The authors use random measurements to measure the second-order Rényi entropy of subsystems \cite{brydges2019probing}. \emph{Demand 2}. Detecting entanglement dynamics beyond the measurement time. Non-equilibrium quantum simulations \cite{abanin2019colloquium,kaufman2016quantum,altman2018many}, such as quantum thermalization, typically require long-time dynamics, but current NISQ devices still have limited coherence times. This implies that measuring long-time entanglement will be difficult. Several studies have used neural networks to investigate the long-time dynamics of local observables beyond the measurement time \cite{mazza2021machine}. But it remains an open question whether it is possible to predict long-time entanglement at an unseen future time based on local measurement data in an observable time window.

   \begin{figure*}
    \centering
    \includegraphics[width=0.65\textwidth]{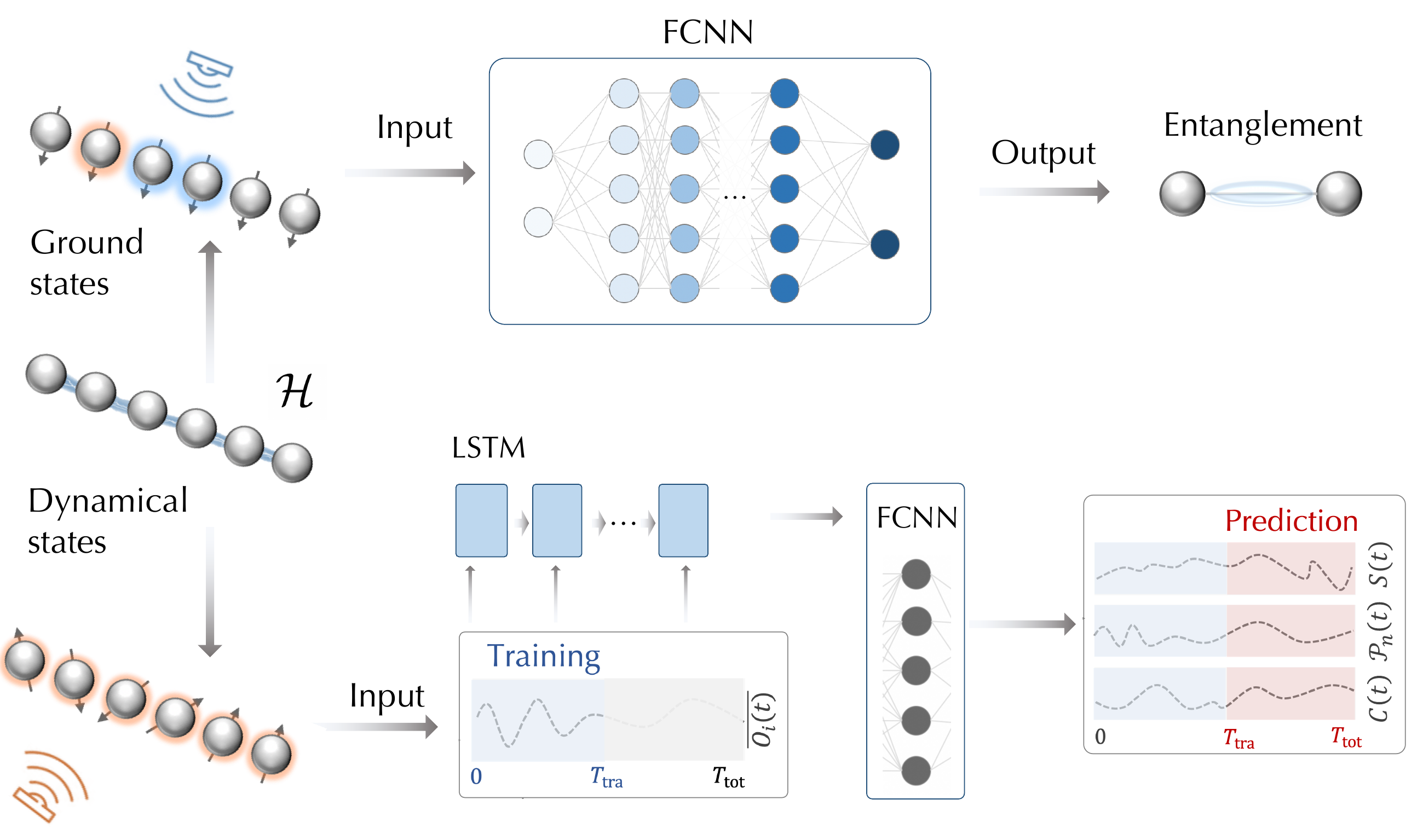}
    \caption{Schematic workflow of our neural network for learning entanglement $\mathcal{E}\equiv\{\mathcal{S}^{(n)}, \mathcal{P}_n, \mathcal{C}\}$ from local Pauli measurements. (a) For the ground states $\ket{\psi_g}$, local Pauli operators are measured and they are directly used to learn $\mathcal{E}$ via FCNN. (b) For the dynamical states $\ket{\psi_t}$, we only measure and input the expectation values of single-qubit Pauli operators in  the range $[0, T_{\text{tra}}]$. It can predict not only the dynamics of $\mathcal{E}$ during the training window, but also the long-time dynamics of $\mathcal{E}$ at the unseen time $[T_{\text{tra}}, T_{\text{tot}}]$.}
    \label{fig1}
\end{figure*}

In this work, we propose a machine learning-assisted detection protocol to determine entanglement from local measurements and validate it by quantifying the entanglement of the ground and dynamical states of the local Hamiltonian. For the ground state, the PT moments and entanglement entropies can only be estimated from two-local Pauli measurements. For dynamical states, the entanglement dynamics beyond the measurement time are also accurately predicted from the single-qubit time traces only in the previous time window. Moreover, we implement an experiment to demonstrate the feasibility of our approach. We employ it to measure static entanglement in ground states and entanglement dynamics in dynamical states of a one-dimensional spin chain on the nuclear magnetic resonance (NMR) platform. QPTs are characterized by measuring the static entanglement as a function of system parameters, and two-local measurements provide accurate predictions of static entanglement as a function of system parameters. For the dynamical case, the entanglement dynamics beyond the measurement time are accurately estimated from the single-qubit time traces using our machine learning approach. Our approach offers a wide variety of applications in the study of quantum many-body physics, from the detection of quantum phase transitions to fascinating non-equilibrium phenomena such as thermalization.

%%%%%%%%%%%%%%%%%%%
\textit{Protocol.} --
The characterization and measurement of quantum entanglement is a crucial task in quantum simulation \cite{RevModPhys.86.153,amico2008entanglement}, which typically prepares the ground state $\mathcal{H}\ket{\psi_g}=E_g\ket{\psi_g}$ or realizes the dynamical state $\ket{\psi (t)}=\text{exp}(-i\mathcal{H}t)\ket{\psi_0}$ of a given local Hamiltonian $\mathcal{H}=\sum_{i=1}^mc_iB_i $ with Pauli basis $B_i$. Their entanglement can be characterized by measuring the entanglement entropy and PT moment. The entanglement entropy provides information about the entanglement contained in the system. It is calculated using the equation $\mathcal{S}^{(n)}=\frac{1}{1-n}\text{log}(\text{tr}\rho_A^n)$.
$\rho_A$ is the reduced density matrix of the total system $\rho_{AB}$. In particular, $\mathcal{S}^{(2)}$ is the second-order Rényi entropy, which has been used to study entanglement growth and thermalization. The PT moment is defined as $\mathcal{P}_n=\text{Tr}[(\rho_{AB}^{T_A})^n]$. 
$\rho_{AB}^{T_A}$ is a partial transpose with respect to subsystem $A$. $\mathcal{P}_1=1$, $\mathcal{P}_2=\text{Tr}[\rho_{AB}^2]$, and $\mathcal{P}_3$ is the lowest PT moment that provides information about the entanglement.
 The first three PT moments have been used to test the bipartite entanglement \cite{elben2020mixed}. Quantum coherence in many-body simulation embodies the essence of entanglement in the following form of $\mathcal{C}=S(\rho_{\text{diag}})-S(\rho)$ \cite{Baumgratz2014,Xi_Li_Fan_2015}. $\rho_{\text{diag}}$ is the diagonal matrix obtained by removing all off-diagonal elements from $\rho$. The above set of entanglement quantities $\mathcal{E}\equiv\{\mathcal{S}^{(n)}, \mathcal{P}_n, \mathcal{C}\}$ commonly requires FST or multi-copy measurements.

 To avoid these issues, we use machine learning to directly predict $\mathcal{E}$ from local Pauli measurements $\mathcal{O}$ on the ground state $\ket{\psi_g}$ or dynamical state $\ket{\psi (t)}$ of the local Hamiltonian $\mathcal{H}$. Figure \ref{fig1} presents the principle of our neural network. The nonlinear relationship between $\mathcal{O}$ and $\mathcal{E}$ can be approximated by a multi-layer neural network with a finite number of neurons. This method train a neural network with a large set of known inputs $\mathcal{O}$ and outputs $\mathcal{E}$. Once the models are trained to convergence, they can be used to experimentally predict the unknown $\mathcal{E}$ from the measured $\mathcal{O}$, without quantum FST. For the ground state, the input $\mathcal{O}=\{\bra{\psi_g}B_i\ket{\psi_g}, 1\leqslant i\leqslant m\}$ is the set of the expectation values of local Pauli operators $B_i$ on the ground state $\ket{\psi_g}$ and the output is static entanglement  $\mathcal{E}$ of $\ket{\psi_g}$. Here, a fully-connected neural network (FCNN) is employed to map the relationship between input $\mathcal{O}$ and output $\mathcal{E}$. For the dynamical state, the input $\mathcal{O}=\{ \bra{\psi_{s\tau}}\sigma_{x,y,z}^{(i)}\ket{\psi_{s\tau}},  1 \le s \le S, 1 \le i \le N\}$ is the set of the expectation values of single-qubit Pauli operators of each qubit at each moment $s\tau$. $S$ is the number of sampling points, and $\tau=T_{\text{tra}}/S$ is the sampling interval. The measured data at the moment $s \tau$ is fed into the $s$-th long short-term memory (LSTM) cell before using FCNN to decode the data. The dynamical entanglement $\mathcal{E}(t)$ is the output result. More intriguingly, once trained, the trained model is able to predict the long-time entanglement $\mathcal{E}(t)$ in an unseen time window $[T_{\text{tra}}, T_{\text{tot}}]$ based on the measurement $\mathcal{O}(t)$ in $[0, T_{\text{tra}}]$. This means that our neural network is able to measure the entanglement dynamics beyond the measurement time. We train the neural network using adaptive moment estimation, a well-known optimizer in machine learning. More details about our neural network can be found in \cite{sm}.

\begin{figure}
    \centering
    \includegraphics[width=0.48\textwidth]{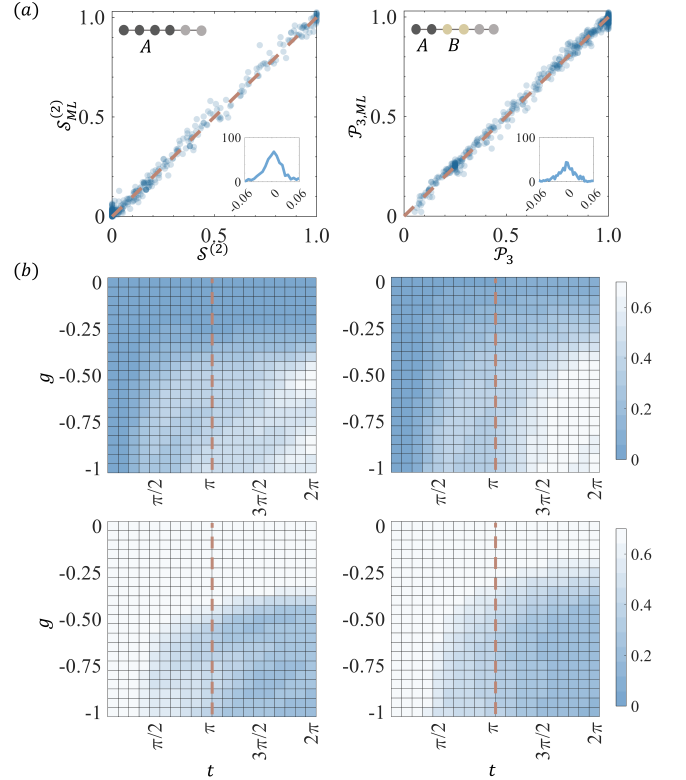}
    \caption{The entanglement estimated by machine learning. (a) The correlation figures between the predicted entanglement $\mathcal{E}_{\text{ML}}$ from 2-local measurements and the theoretical $\mathcal{E}$. The inset figures are the distributions of the difference  $\mathcal{E}_{\text{ML}}-\mathcal{E}$. (b) Prediction of entanglement dynamics from single-qubit time traces $\mathcal{E}_{\text{ML}}(t)$. The right column is the prediction result obtained by our machine learning method and the left column is the theoretical values. The input layer contains only the measured single-qubit time traces in $[0,\pi]$. The trained model allows us to predict  $\mathcal{E}_{\text{ML}}(t)$ at the unseen time $[\pi,2\pi]$. The measured subsystems are represented by the gray rounded schematic.}
    \label{fig2}
\end{figure}

%%%%%%%%%%%%%%%%%%%%%%%%%%%%%%%%%%%%%%%%%%%%%%%%%%%%%
\textit{Numerical results.} -- To demonstrate the feasibility of our neural network, we numerically test the following Hamiltonian models. \emph{Model 1.} We consider a 6-qubit 2-local Hamiltonian $\mathcal{H}=\sum_{i=1}^6\bm{\omega}^i \cdot \bm{\sigma}^i+\sum_{j=1}^5\bm{\sigma}^j\cdot \bm{J}^j \cdot \bm{\sigma}^{j+1}$ and train a neural network to predict the static $\mathcal{E}$ of the ground states. $\bm{\sigma}^i=(\sigma_x^i, \sigma_y^i, \sigma_z^i)$ is the vector of Pauli matrices. $\bm{\omega}^i=(\omega_x^i, \omega_y^i, \omega_z^i)$ and $\bm{J}^j=(\bm{J}_{xx}^j, ..., \bm{J}_{zz}^j)$ represent the external magnetic field strength and the coupling tensor, respectively. During training, we generate a large number of ground states $\ket{\psi_g}$ by randomly choosing $\bm{\omega}, \bm{J}\in [-1, 1]$, and using $\mathcal{O}$ and $\mathcal{E}$ of the ground states as input and output, respectively. $\mathcal{O}$ is the set of the expectation values of the one and two-body Pauli measurements of the state, which can be easily obtained. The predicted $\mathcal{E}$ includes the entanglement information of the subsystems. We generate 100,000 pairs of such ($\mathcal{O}$, $\mathcal{E}$) for training the neural network and then randomly select 500 for testing the performance. In Fig. \ref{fig2}(a), we compare the estimation of $\mathcal{S}^{(2)}$ (the subsystem $A=1234$) and $\mathcal{P}_3$ (the subsystems $A=12$ and $ B=34$) by our neural network with the traditional FST. \emph{Model 2.} We train a neural network capable of predicting the entanglement dynamics of the non-equilibrium states. The system starts from $\ket{\psi_0}=R_z(\pi/8)R_y(\pi/8)\ket{0}^{\otimes 6}$ and evolves into $\ket{\psi_t}=\text{exp}(-i\mathcal{H}_{\text{d}}t)\ket{\psi_0}$ under the Hamiltonian $\mathcal{H}_{\text{d}}$. $\mathcal{H}_{\text{d}}$ is defined as $\mathcal{H}_{\text{d}}=J\sum_{i=1}^5\sigma_z^i\sigma_z^{i+1}+g\sum_{j=1}^6 \sigma_x^i$, where $J$ and $g$ are adjustable parameters. $\mathcal{H}_{\text{d}}$ is one of the models used in the study of dynamical quantum phase transitions (DQPTs)  \cite{Nie2020experimental}. We generate various $\mathcal{H}_{\text{d}}$ by randomly selecting $J$ and $g$ between $-1$ to 1, allowing the initial state to evolve in various ways to a large number of $\ket{\psi_t}$. The neural network is trained by taking ($\mathcal{O}$, $\mathcal{E}$) of each state at one time as input and output.
 This model differs from the previous one in that we use only single-qubit Pauli measurements. The predicted $\mathcal{E}$ includes the entanglement dynamics of the subsystems. Here, we still use 100,000 data for training the neural network, and we only feed $\mathcal{O}$ in the previous time $[0, \pi]$ as the input during training. The output is the entanglement dynamics $\mathcal{E}(t)$ in a longer time range $[0, 2\pi]$. We set $J=-0.5$ and divide $g$ from $-1$ to $0$ into 20 parts to generate 20 pieces of data to test our neural network. Figure \ref{fig2}(b) depicts the predicted $\mathcal{S}^{(2)}(t)$ (the subsystem $A=1$) and $\mathcal{P}_3(t)$ (the subsystems $A=1$ and  $B=23$) for the time interval $[0, 2\pi]$. It is shown that $\mathcal{E}(t)$ can be accurately predicted from the single-qubit time traces and that DQPTs are revealed across $g=-0.5$ \cite{Nie2020experimental}. More machine learning results and the training details can be found in \cite{sm}.

%When we test, we input $\mathcal{O}$ in $[0, \pi]$ time, and the neural network predicts $\mathcal{O}$ in $[0, 2\pi]$ time. Let 

%%%%%%%%%%%%%%%%%%%%%%%%%%%%%%%%%%%%%%%%%%%%%%%%%%%%%
\textit{Experiment.} -- We also adopt our neural networks to detect equilibrium and dynamical quantum phase transitions on a 4-qubit nuclear magnetic resonance (NMR) platform \cite{jones2011quantum,gershenfeld1997bulk}. The used four-qubit sample is $^{13}$C-labeled trans-crotonic acid dissolved in $d$6-acetone, where four $^{13}$C nuclear spins are encoded as a 4-qubit quantum processor. The internal Hamiltonian of the system is given by  
\begin{equation}
\mathcal{H}_{\text{int}}=-\sum_{i=1}^4\pi\nu_i\sigma_z^i+\sum_{i,j}^4\pi\frac{J_{ij}}{2}\sigma_z^i\sigma_z^j.
\end{equation} 
$\nu_i$ is the chemical shift of each spin, and $J_{ij}$ is the coupling strength between different spins. The spin dynamics is controlled by shaped radio-frequency (rf) pulses \cite{vandersypen2005nmr}. The molecular structure and the Hamiltonian parameters can be found in the supplemental material\cite{sm}. All experiments were carried out on a Bruker 600-MHz spectrometer at room temperature. 

First, we observe equilibrium QPTs in two types of spin-half chains by studying the entanglement of their ground states. Their Hamiltonians are defined as $ \mathcal{H}_{\text{XXZ}}=-J\sum_{i=1}^{3}(\sigma_x^i\sigma_x^{i+1}+\sigma_y^i\sigma_y^{i+1})+\Delta\sum_{j=1}^{3}\sigma_z^i\sigma_z^{i+1}$ and $\mathcal{H}_{\text{XX}}=-J\sum_{i=1}^{3}(\sigma_x^i\sigma_x^{i+1}+\sigma_y^i\sigma_y^{i+1})+h_z\sum_{j=1}^{4}\sigma_z^i$. Their ground states exhibit QPTs characterized by sudden entanglement as a function of $\Delta$ or $h_z$ \cite{Son_2009,Zhang_2011}. $\Delta$ is the anisotropic parameter characterizing the magnetic field. In experiments, we prepare 50 ground states by changing $\Delta$ and $h_z$ from $-1$ to 1 with a step of 0.04, measure the expectation values of the two-local Pauli operators, and then use our trained neural network in Model 1 to predict the entanglement information of these states. Second, we investigate the non-equilibrium phenomena by characterizing the entanglement evolution $\mathcal{E}(t)$ of the dynamical states of $\mathcal{H}_{\text{d}}$. Experimentally, we prepare the initial state $\ket{\psi_0}=R_z(\pi/8)R_y(\pi/8)\ket{0}^{\otimes 4}$ and implement the dynamical evolution of the two Hamiltonians with the parameters $J=-0.5,g=-0.3$ and $J=-0.5,g=-0.75$. We then measure the single-qubit time traces and use the trained neural network to predict the entanglement dynamics $\mathcal{E}(t)$. Here we also perform quantum FST on the ground states and dynamical states to provide a comparison with machine learning results.

\begin{figure}
%\begin{figure}
    \centering
    \includegraphics[width=0.5\textwidth]{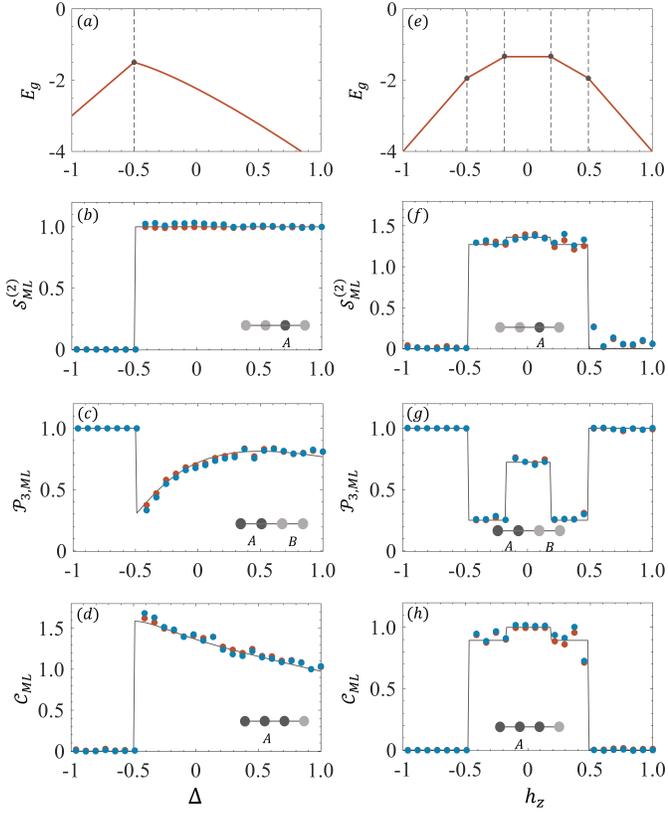}
    \caption{Phase diagram of $\mathcal{H}_{\text{XXZ}}$ and $\mathcal{H}_{\text{XX}}$ models. (a) and (e) are the ground energy levels (red lines). (b-d) and (f-h) show the entanglement $\mathcal{E}$ of the ground states of $\mathcal{H}_{\text{XXZ}}$ and $\mathcal{H}_{\text{XX}}$ models, respectively. Theoretical calculation (solid lines), quantum FST (red points), and our neural networks results (blue points) are distinguished. The measured subsystems are represented by the gray rounds schematic.}
    \label{fig3}
%\end{figure}
\end{figure}

\begin{figure}[tbh]
  %\begin{figure}
      \centering
      \includegraphics[width=0.45\textwidth]{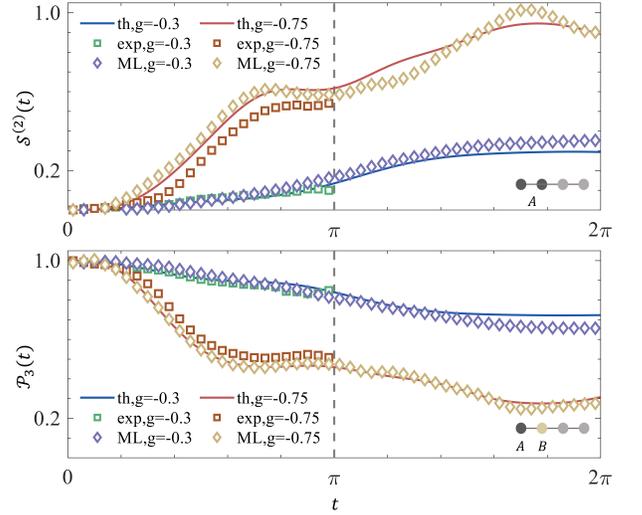}
      \caption{The dynamical evolution $\mathcal{S}^{(2)}(t)$ (the subsystem $A=12)$,  $\mathcal{P}_3(t)$ (the subsystems $A=1, B=2)$ for two sets of parameters (i) $J=-0.5, g=-0.3$ and (ii) $J=-0.5, g =-0.75$. The input layer only contains the measured single-qubit time traces in $[0, \pi]$. The trained model allows us to predict $\mathcal{E}(t)$ in the training window $[0, \pi]$ and in the unseen future time window $[\pi, 2\pi]$. The theoretical results (solid lines), quantum FST results (square dots), and predicted results (diamond dots), for the two setups, are each represented by a different color of identification line (or dot).}
      \label{fig4}
  %\end{figure}
  \end{figure}

Our experiment consists of the following steps. \emph{(i) Initialization}. We first initialize the spins to pseudo-pure state (PPS) $\ket{0}^{\otimes N}$ from the highly-mixed state via the selective-transition method \cite{cory1997ensemble,Lu2011SimChemiNMR,Lu_2014}. Quantum FST was also performed to check the PPS quality. A fidelity of more than 99\% provides a reliable initialization for the following experiments. \emph{(ii) Preparing target states.} For the ground states of $\mathcal{H}_{\text{XX}}$ and $\mathcal{H}_{\text{XXZ}}$, we optimize a 20 ms shaped pulse to drive the system to the target states from $\ket{0}^{\otimes 4}$. For the dynamical states of $\mathcal{H}_{\text{d}}$, we also optimize a 30 ms shaped pulse that realize the evolution $\text{exp}(-i\mathcal{H}_{\text{d}}t)$. All shaped pulses are searched with the gradient ascent pulse engineering (GRAPE) technique \cite{khaneja2005optimal}.  \emph{(iii) Measuring local information}. We measure the expectation values of $\mathcal{O}$. It consists of 39 Pauli measurements (3$\times$4 single-qubit Pauli operators $\{\bm{\sigma}^i\}$, 3$\times$9 two-qubit Pauli operators $\{\bm{\sigma}^i\bm{\sigma}^{i+1}\}$) for the static models $\mathcal{H}_{\text{XXZ}}$ and $\mathcal{H}_{\text{XX}}$ \cite{sm}. For the dynamical states of $\mathcal{H}_{\text{d}}$, we measure 50 temporal points from 0 to $\pi$ with a step of $\pi/50$ and measure $3\times 4$ single-qubit Pauli operators $\{\bm{\sigma}^i\}$ each moment.  As an ensemble system, NMR can easily measure the expectation value of the Pauli operators. \emph{(iv) Predicting entanglement}. We first trained the neural networks with 100,000 training data for both the static and dynamic models. We then feed the measured data in the above step into the trained neural network to predict the entanglement of the target state.

Our neural networks can reveal both static QPTs and DQPTs that are consistent with data from quantum FST and theoretical ways. In Fig. \ref{fig3}, we show the static entanglement obtained by theoretical calculation (solid line), quantum FST (square dots), and our neural networks (diamond pots). In the $\mathcal{H}_{\text{XXZ}}$ model, we set $J=-0.5$, and the first-order QPT occurs when the model reverts to an isotropic Heisenberg model ($\Delta=J$). In the $\mathcal{H}_{\text{XX}}$ model, we set $J=-0.3$, and we can observe QPT occurs at the magnetic critical point where the ground state energy levels cross ($h_z= 2 J \cos ( \frac{k \pi}{5} ) $, where $1 \leq k \le 4$ is an integer) \cite{PhysRevA.79.022302}. In Fig. \ref{fig4}, we show the dynamical nature of  the entanglement $\mathcal{E}(t)$ in $[0,2\pi]$ using the measured single-qubit time traces in $[0,\pi]$. Since the phase transition point for non-equilibrium DQPTs is $g_c=-0.5$ \cite{Nie2020experimental}, there will be different dynamic behaviours in the cases of $g < g_c$ and $g > g_c$. When $g=-0.3$, the oscillation amplitude of $\mathcal{E}(t)$ (blue line) is modest and close to its initial value, and when $g=-0.75$, $\mathcal{E}(t)$ (red line) oscillates range is greatly larger \cite{sm} .

%%%%%%%%%%%%%%%%%%%%%%%%%%%%%%%%%%%%%%%%%%%%%%%%%%%%%
\textit{Conclusion.} -- In summary, we have designed a machine learning-assisted strategy to estimate entanglement information from local measurements for both ground and dynamical states. It can skip the challenging quantum FST, which in general requires an exponential number of measurements. At the same time, for dynamical states, neural networks can keep track of long-term entanglement for an unseen future time. Our numerical simulations show that our neural network is able to estimate the integrated entanglement information with significant accuracy. We have furthermore verified the feasibility of our approach in practical 4-qubit NMR experiments. Experimental results have shown that the entanglement in the ground and dynamical states of special 4-qubit Hamiltonians can also be accurately estimated, and QPTs and DQPTs have been observed by characterizing the entanglement properties. 

Our neural network strategy has the following extensions and future applications. First, we can extend our idea to measure the entanglement of different quantum states. Previous research has demonstrated that some quantum states can be determined using compressing sensing \cite{donoho2006compressed,flammia2012quantum}, direct estimation \cite{flammia2011direct}, and the UD-property (Uniquely Determined, UD) \cite{xin2017quantum}. These techniques normally measure random or fixed Pauli measurements. This means that our framework can be extended to these types of quantum states. Second, we do not consider noise in the present framework. The experimental validity of our approach also supports its applicability in practical quantum devices. In the future, we can test and improve the robustness by incorporating noise into the training data. Third, our framework will have wide applications in studying the intriguing equilibrium and non-equilibrium phenomena \cite{Hinrichsen_2000}, because the entanglement entropy is an essential quantify that diagnoses and characterizes quantum phase transitions \cite{PhysRevA.71.012307}, quantum thermalization \cite{Kaufman_2016}, and quantum MBL \cite{Rispoli_2019}.

\begin{acknowledgments}
\textit{Acknowledgments.} -- This work is supported by the National Natural Science Foundation of China (12275117, 11905099, 12075110, 11975117, 11875159  and U1801661), Guangdong Basic and Applied Basic Research Foundation (2022B1515020074, 2019A1515011383, 2021B1515020070), National Key Research and Development Program of China (2019YFA0308100), Guangdong International Collaboration Program (2020A0505100001), Science, Technology and Innovation Commission of Shenzhen Municipality (ZDSYS20170303165926217, KQTD20190929173815000, JCYJ20200109140803865, JCYJ20170412152620376 and JCYJ20180302174036418),  Shenzhen Science and Technology Program (Grants No. RCYX20200714114522109 and No. KQTD20200820113010023), Pengcheng Scholars, Guangdong Innovative and Entrepreneurial Research Team Program (2019ZT08C044), and Guangdong Provincial Key Laboratory (2019B121203002). Y. H. and L. C. contributed equally to this work. \end{acknowledgments}

%\bibliography{LearnEnt.bib}

%

\end{document}